\documentstyle[preprint,prl,aps,floats,epsf]{revtex}

\tightenlines

\newcommand{\beq}{\begin{equation}}
\newcommand{\eeq}{\end{equation}}
\newcommand{\be}{\begin{eqnarray}}
\newcommand{\ee}{\end{eqnarray}}
\newcommand{\ben}{\begin{eqnarray*}}
\newcommand{\een}{\end{eqnarray*}}

\begin{document}

\draft

\title{Lifetime of a Disoriented Chiral Condensate}

\author{James V. Steele$^1$ and Volker Koch$^2$}
\address{$^1$Department of Physics,
The Ohio State University,\ Columbus, OH\ \ 43210 \\
$^2$Lawrence Berkeley National Laboratory, Berkeley, CA\ \ 94720}

\date{\today}

\maketitle

\begin{abstract}
The lifetime of a disoriented chiral condensate formed within a
heat bath of pions is calculated assuming temperatures and densities
attainable at present and future heavy-ion colliders.  A
generalization of the reduction formula to include coherent states
allows us to derive a formula for the decay rate.  We predict 
the half-life to be between 4 and 7~fm/c, 
depending on the assumed pion density.
We also calculate the lifetime in the presence of higher resonances and
baryons, which shortens the lifetime by at most 20$\%$.
\end{abstract}

\thispagestyle{empty}%

\newpage%


Formation of hot and dense matter in heavy-ion collisions has the
possibility of creating a phase where chiral symmetry is restored.
As this matter cools and expands, the vacuum could relax into the
``wrong'' zero temperature ground state.  
Subsequent shifting of the vacuum back into alignment with the outside
world could then lead to an excess of low momentum
pions \cite{raja} in a single direction in isospin space \cite{kowal}. 
This excess is called a disoriented chiral condensate (DCC) and
many studies have looked into whether its presence could be a signal
of chiral symmetry restoration in heavy-ion collisions.  

Most of the
discussion has centered around the details of formation of such
regions~\cite{raja,kowal,All,orig,Greiner,Rischke}. 
There has been a general consensus that these
regions could be numerous and large enough with the conditions at RHIC
and the LHC to be detected. 
While the idea of DCC-formation is appealing,
the physics governing the chiral phase transition 
is not known well enough to make reliable predictions
about the possible formation of these condensates.
Experiment is needed to establish their existence.

The ability to detect a DCC from hadronic observables in heavy-ion
collisions depends on the condensate lifetime.
An early estimate~\cite{orig} gave a half-life of 
$\tau\sim 3$ fm/c. 
A more recent
calculation~\cite{Greiner} has found a damping rate of $\gamma \sim 1$
(fm/c)$^{-1}$. 
This roughly corresponds to a half-life of
$\tau\sim1/\gamma\sim1$ fm/c, which would be short enough to
jeopardize the definiteness of the signal.  This calculation was based
on the $O(4)$ sigma model in the symmetric phase.  
However, after
formation, the DCC lives in the phase of spontaneously broken
chiral symmetry.  Characteristic of this phase is the suppression of $S$-wave
scattering among pions, which should protect the low momentum pion
modes in the DCC.  Indeed, an extension of Ref.~\cite{Greiner} to
the broken phase produces smaller damping rates~\cite{Rischke}.

The purpose of this letter is to provide a reliable estimate of the
lifetime of a 
DCC state in the hadronic (chirally broken) phase. Defining the DCC to
be a coherent state of pions with low momenta, we
derive a  general formula for the decay rate in the presence of other
hadrons. The effect of higher resonances as well as that of baryons to
the DCC lifetime is also estimated. 
Our calculation is constrained by data at every point possible.

Assuming the formation of a DCC, interactions
with the thermal heat bath can enhance or deplete the number of pions
in the condensate.  In a heavy-ion collision, pions are the most
abundant thermal particles, and so their interactions with the
DCC is expected to give the dominant contribution to the decay
rate.  The DCC can be written as a coherent state~\cite{kowal,orig}
\beq
| \eta \rangle = e^{-N/2}
e^{\int\! d{\tilde k}\; \eta_k a_k^\dagger} |0 \rangle\, ,
\eeq
normalized to unity and for definiteness taken to consist of neutral
pions, $\pi^0$, created by $a_k^\dagger$.  The Lorentz invariant
measure $d{\tilde 
k}=d^3k/(2\pi)^3 2 E_k$ is weighted by the normalizable function
$\eta_k$ to give a momentum distribution which could evolve
over time \cite{IZ}.  The number of particles in the DCC is then 
given by $N=\int\! d{\tilde k}\; |\eta_k|^2$.

We can get an estimate of the decay rate by considering two
body interactions of the DCC pions with the surrounding thermal pions.
Contributions come from $\pi\pi$ scattering with
either two thermal pions knocking one pion into the condensate or a
DCC pion interacting with a thermal pion to escape from the 
condensate. There could also be individual scatterings with two pions
knocked out or put into the condensate, but these should be 
suppressed due to the restricted phase space of the DCC.  Using
\beq
\langle \eta | \pi^0(x) | \eta \rangle = 
\int\! d{\tilde k}\; e^{-i k x} \langle \eta | a_k | \eta \rangle =
\int\! d{\tilde k}\; e^{-i k x} \eta_k \, ,
\eeq
a rederivation of the Lehmann-Symanzik-Zimmerman (LSZ) reduction formula
taking into account the coherent state gives
\be
\frac{dN}{dt} &=& \int\! d{\tilde k}_1\, d{\tilde k}_2\,
d{\tilde k}_3\, 
F_{123}\,  
\langle | {\cal T}_{\pi\pi} |^2\rangle 
\nonumber\\
&&\times 2\pi\, \delta (E_0+E_1-E_2-E_3) \,
\frac{|\eta_{k_2+k_3-k_1}|^2}{(2E_0)^2}  \, .
\label{prerate}
\ee
The matrix element for $\pi\pi$ scattering 
$\langle |{\cal T}_{\pi\pi}|^2\rangle=\frac12 \sum_I \frac13 (2I+1) 
|T^I_{\pi\pi}|^2$ is isospin averaged to account 
for only $\pi^0$'s coming from the DCC and the factor of $\frac12$ 
accounts for identical particles in the final state after isospin
considerations. The thermal weighting is given by 
\beq
F_{123}=f_2 f_3(1+f_1)-f_1 (1+f_2)(1+f_3) ,
\eeq
with $f_i=(\exp(E_i/T)-1)^{-1}$ representing the Bose-Einstein
momentum distribution.  
The DCC pion energy is $E_0=[( k_2+ k_3- k_1)^2+m_\pi^2]^{1/2}$.

The phase space correlations in Eq.~(\ref{prerate}) can be simplified
for DCC pions narrowly peaked near a single momentum $k_0$.  
Then the square of the momentum distribution 
can be replaced by a momentum-conserving delta-function
$|\eta_{k_2+k_3-k_1}|^2\to N (2\pi)^3\, 2 E_0\,
\delta^3(k_0+k_1-k_2-k_3)$,
giving the decay rate
\beq
\!\!\!\frac1{N} \frac{dN}{dt}\!\! =\! \frac1{2E_0}\! 
\int\!\! d{\tilde k}_1\,
d{\tilde k}_2\, d{\tilde k}_3\, 
F_{123}\, \langle |{\cal T}_{\pi\pi}|^2\rangle \,
(2\pi)^4 \delta^4(\Sigma\, k_i) 
\, . 
\label{rate}
\eeq
Although this formula pertains to two-particle interactions,
it can easily be generalized to include more particles.  
To understand the various contributions to the decay rate,
we first take the DCC pions to be at rest, $k_0=(m_\pi, {\bf 0})$.
We relax this condition in Eq.~(\ref{rate2}), allowing for a finite
spread in the momenta in accordance with Eq.~(\ref{prerate}).

This result was derived assuming the removal or addition of a pion to
the condensate does not effect the DCC coherent state.
This approximation is valid to better than 10\% until the number of
particles in the condensate decreases below $N\sim 5$, at which point
the existence of a macroscopic condensate is questionable anyway.
Eq.~(\ref{rate}) can also be derived starting from the Boltzmann
equation by assuming the DCC momentum distribution function obeys
$f_0\sim 1+f_0$, which again is appropriate for large enough $N$ and
distributions peaked at zero momentum.

We can now estimate the decay rate by replacing the $\pi\pi$
scattering matrix element by its experimental value in terms of the
center of mass energy $\sqrt{s}$, three-momentum $q$, and scattering
angle $\theta$, 
\beq
T^I_{\pi\pi}= 32\pi \sum_l\, (2l+1) P_l(\cos \theta)\,
\frac{\sqrt{s}}{2q}\, e^{i\delta^I_l}\sin\delta^I_l\, .
\eeq
Accounting for the three dominant phase shifts $\delta_0^0$,
$\delta_1^1$, and $\delta_0^2$ by using a parameterization of the data
\cite{Bertsch}, the momentum integral in the decay rate
can be readily evaluated.  Defining the half-life $\tau$ by
\beq
\frac1{N} \frac{dN}{dt} \equiv - \frac1{\tau} ,
\eeq
we find $\tau=8.9$~fm/c for $T=150$~MeV and $\tau=5.6$~fm/c for
$T=170$~MeV.  Considering the entire hadronic phase exists for about
$10-20$~fm/c, the DCC could live for a substantial
fraction of the hadronic phase, allowing for a strong signal
of its existence, {\it e.g.}~in the dilepton channel \cite{dilep}. 
Even assuming three times the number of thermal
pions, as predicted by some event generators \cite{rqmd}, the
half-life is still $\tau=3.8$ fm/c at $T=150$ MeV.  

It should be noted that the momentum dependence of the phase shifts is
important to the result.  Approximating the $\pi\pi$ amplitude
purely by threshold ($q=0$) scattering lengths gives a half-life five 
times longer.  Only if the range terms, which have a $q^2$ dependence,
are also kept does the estimate come within $10\%$ of the full phase shift
result. 
The $\rho$ channel ($P$-wave) 
contribution is merely a few percent effect, in agreement with
estimates using a simple Breit-Wigner form.
The main contribution to
the decay rate comes from the $S$-wave.  This is
because the other phase shifts are very small in the region where the
Bose-Einstein distribution functions are appreciable.  

\begin{figure}
\begin{center}
\leavevmode
\epsfxsize=4in
\epsffile{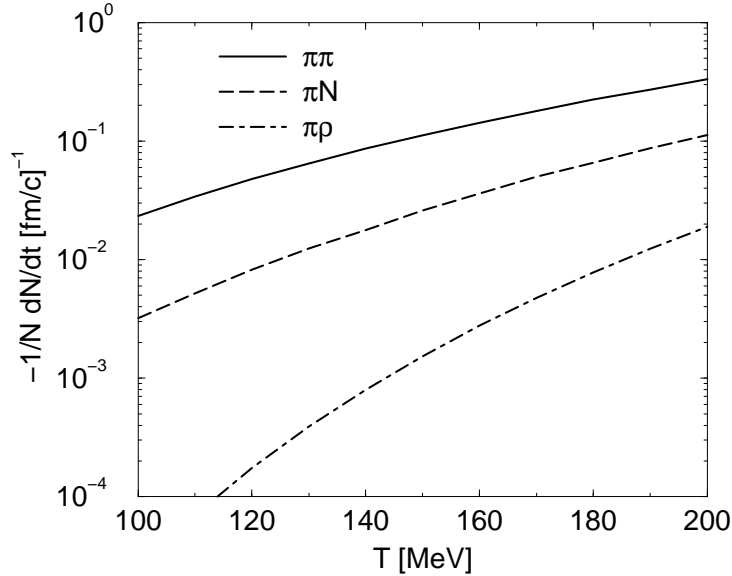}
\end{center}
\caption{\label{decay} The contribution to the decay rate,
Eq.~(\protect\ref{rate}), in inverse fm/c from $\pi\pi$, $\pi\rho$,
and $\pi N$ scattering keeping the ratio of pions to nucleons fixed at five.}
\end{figure}

The $\rho$ resonance could possibly play a role as a thermal particle
though.  We therefore look at $\pi\rho$ scattering.  The dominant
contribution comes from the formation of an $a_1(1240)$ meson in the
$s$-channel, which we can model by a relativistic $S$-wave
Breit-Wigner matrix element
\beq
\langle |{\cal T}_{\pi\rho}|^2 \rangle = \frac{(8\pi s)^2}{q^2}
\frac{3\,\Gamma_{a1}^2(s)}{(s-m_{a1}^2)^2+s\Gamma_{a1}^2(s)} 
\eeq
with center of mass three-momentum $q$ and a momentum dependent width
whose form is dictated by general decay of an axial-vector particle
into a vector and pseudoscalar particle ($q_{a1}=q|_{s=m_{a1}^2}$) 
\beq
\Gamma_{a1}(s)= 400\; \frac{q\,(3m_\rho^2+q^2)}{q_{a1}(3m_\rho^2+q_{a1}^2)}
\, \mbox{MeV}\, .
\eeq
Adding this matrix element to Eq.~(\ref{rate}) gives a negligible
effect, not changing the original $\pi\pi$ scattering half-life 
to the accuracy quoted.  
This can be seen in Fig.~\ref{decay}, where the matrix
elements for $\pi\pi$ and $\pi\rho$ scattering are compared for various
temperatures.

The Boltzmann suppression of the $a_1$ pole along with threshold
suppression from the width conspire to make $\pi\rho$ scattering a
small effect. Other meson-meson processes not considered here, such as 
$\pi a_1$ scattering, are therefore expected to be smaller.

Baryons could also play a substantial role in reducing the lifetime of a DCC.
Even at SPS energies ($E_{cm} = 20 \, \rm GeV/nucleon$) a large number of 
baryons are found in the central rapidity region.
We therefore study the
dependence of the half-life on nucleon density.
The matrix element for $\pi N$ scattering is \cite{weise}
\beq
\langle |{\cal T}_{\pi N}|^2 \rangle = 64 \pi^2 s
\sum_I \frac13 (2I+1) \left(|g_I(s,\theta)|^2 + |h_I(s,\theta)|^2
\right) 
\eeq
with $g_I(s,\theta)$ and $h_I(s,\theta)$ denoting the spin non-flip and
spin-flip amplitudes for each of the two possible isospins $I=\frac12,
\frac32$.  These can be written in terms
of phase shifts labeled by $\alpha=l,2I,2J$ 
\ben
g_{1/2}(s,\theta) &=& {\cal F}_{S11}
+ \cos\theta\, \left(2{\cal F}_{P13} + {\cal F}_{P11} \right) + \ldots 
\\
h_{1/2}(s,\theta) &=& {\cal F}_{P13}- {\cal F}_{P11} + \ldots 
\een
with similar expressions for the $I=\frac32$ channel by direct
substitution of the isospin indices.  The ellipses refer to higher
partial waves we will not consider here and 
\beq
{\cal F}_\alpha = \left( q\cot\delta_\alpha(q) - iq \right)^{-1} \, .
\eeq
Adding this matrix element to Eq.~(\ref{rate}), we must
also replace the thermal weighting factor of this term by
\beq
F_{123}^N=f_2 f_3^N(1-f_1^N)-f_1^N(1+f_2)(1-f_3^N) 
\eeq
with $f_i^N=(\exp((E_i-\mu)/T)+1)^{-1}$.  A chemical potential $\mu$
has been introduced to enforce a particular nucleon density.

For a parameterization of the $\pi N$ phase shifts, we use the full
relativistically improved $\Delta(1232)$ isobar model~\cite{weise},
which reproduces the experimental phase shifts very well up to the
inelastic threshold.  
Adding this decay rate to those of $\pi\pi$ and $\pi\rho$ scattering 
without a chemical potential gives only a few percent decrease in the
half-life.  But taking
$\mu=260$ MeV at $T=150$ MeV to enforce a $5:1$ ratio between pions
and nucleons as observed at the SPS \cite{piNratio}, we end up with
a half-life of $\tau=7.2$~fm/c.  The effect of the
nucleons is about a $20\%$ reduction in $\tau$.
Still, this effect is small enough to imply experiments already taking
place at the SPS should be capable of detecting
remnants of DCC formation, provided of course that the conditions for DCC
formation are met at these energies.  

A summary of our results is shown in Fig.~\ref{decay}.
At each temperature the chemical potential of the nucleons is chosen
to enforce the $5:1$ ratio between pions and nucleons.  
Since the effect of the nucleons is small, this result also
applies for RHIC energies ($E_{cm} = 200 \, \rm GeV/nucleon$) where
the central rapidity pion to baryon ratio is probably even larger.

\begin{figure}
\begin{center}
\leavevmode
\epsfxsize=4in
\epsffile{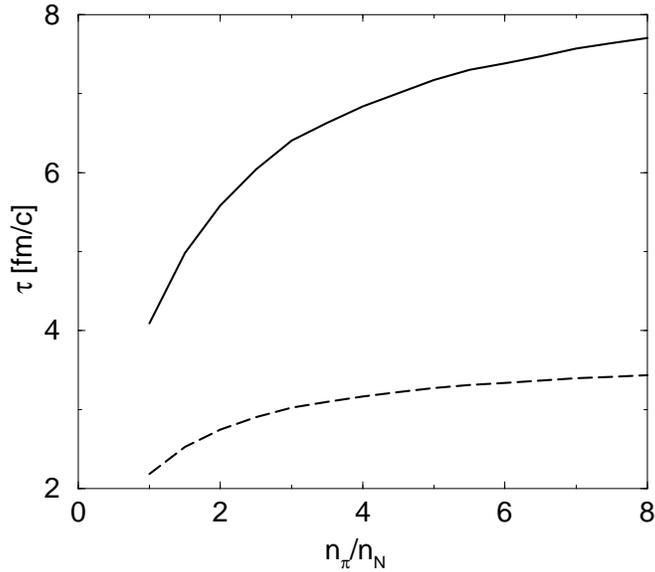}
\end{center}
\caption{\label{dens} The half-life $\tau$ at $T=150$ MeV for various
pion to nucleon ratios. The dashed line is for three times the number
of thermal pions and nucleons.}
\end{figure}

We also show a plot of the dependence of half-life on nucleon density
for $T=150$~MeV in Fig.~\ref{dens}. The $\pi N$ contribution becomes
comparable with the $\pi\pi$ contribution when there is about one
nucleon for every pion.  
However, we have so far assumed completely thermalized pions and
nucleons for this estimate, whereas some event generators indicate 
an enhancement of
the occupation numbers by a factor of three \cite{rqmd}. 
Taking this into account
in our calculation gives the dashed line in Fig.~\ref{dens}, which
shows the half-life is between $3$ and $3.5$~fm/c for all the pion to
nucleon ratios considered.  
If the over-population of particle spectra predicted
by event generators is correct, it implies the DCC signal would be
difficult to detect from hadronic observables, requiring an indirect
means such as dileptons \cite{dilep}.

Rather than taking the condensate to consist of only pions at rest,
we can investigate the change in the lifetime for a finite spread in
the DCC momenta.
This will open up the phase space of the DCC and allow for further
interactions with the heat bath.  
By introducing an integration over the Mandelstam
variables $s=(k_0+k_1)^2$ and $t=(k_0-k_2)^2$ \cite{kapusta}, we can
simplify Eq.~(\ref{rate}) for $\pi P$ scattering 
(with $P$=$\pi,\rho$, or $N$) to 
\be
\frac1N \frac{dN}{dt} = \frac1{(2\pi)^4 16 E_0^3}  \int\!\! ds\, dt\, 
\langle |{\cal T}(s,t)|^2 \rangle
\int\!\! dE_1 dE_2 \, \frac{F_{123}}{\Phi^{1/2}}
\label{rate2}
\ee
with $\Phi=4 K_1^{\ 2} K_2^{\ 2}- \left( K_1^{\ 2}+K_2^{\ 2}-K_3^{\ 2}
\right)^2$, $k_0=(E_0,{\bf k}_0)$,
\ben
K_i^2 &=& -\frac{m_\pi^2}{E_0^2}(E_i^2-\bar{m}_i^2)
-m_i^2 + \frac{E_i}{E_0} q_i^2 - \frac{q_i^4}{4E_0^2} 
\, ,
\\
q_1^2&=&s-m_\pi^2-m_P^2, \quad 
q_2^2=2m_\pi^2-t, \quad 
q_3^2=q_1^2-q_2^2 \, ,
\\
\bar{m}_1^2 &=& m_{1,3}^2 =m_P^2, \quad
\bar{m}_2^2=m_2^2=m_\pi^2, \quad
\bar{m}_3^2=2 E_0 E_3+u \, .
\een
with integration limits dictated by $s>(m_\pi+m_P)^2$, $t<0$,
$K_i^2>0$, $\Phi>0$, and $E_0+E_1>E_2+m_P$.  The 
thermal pion energy should also be restricted to be above the maximum
DCC energy $E_0$.  

\begin{figure}
\begin{center}
\leavevmode
\epsfxsize=4in
\epsffile{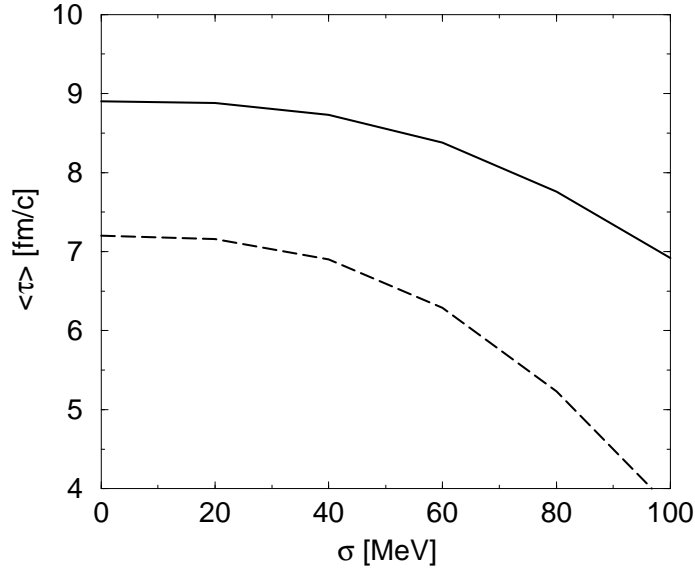}
\end{center}
\caption{\label{width} The average half-life $\langle\tau\rangle$ for
$T=150$ MeV as a function of the width of the DCC pion momentum
distribution $\sigma$ from $\pi\pi$
scattering alone (solid) and also including $\pi\rho$ and
$\pi N$ scattering (dashed).}
\end{figure}

We then average the half-life obtained from Eq.~(\ref{rate2}) over the 
momentum distribution in the DCC as dictated by
Eq.~(\ref{prerate}). 
Taking $|\eta_k|^2\propto\exp(-|{\bf k}_0|^2/\sigma^2)$, 
the average half-life is plotted as a
function of the width $\sigma$ for $T=150$~MeV in Fig.~\ref{width}.
The solid line is for the
$\pi\pi$ scattering contribution alone and the dashed line also
includes the $\pi\rho$ and $\pi N$ contributions.
The pion to nucleon ratio is kept fixed at $5:1$.  The 
decrease of the half-life from $8.9$~fm/c to $7.2$~fm/c for 
$|{\bf k}_0|=0$  was due to the nucleon contribution.  As $\sigma$
increases, 
$P$-wave $\pi\pi$ scattering and, to a lesser extent, 
$\pi\rho$ scattering become more active, decreasing $\langle\tau\rangle$.
Consequently, smaller DCC domains (characterized by a larger $\sigma$)
decay faster.

In conclusion, we have derived a simple decay rate for a DCC and
applied it assuming initial conditions at present and future heavy-ion
colliders.  In all stages of the calculation we have used input from
data to constrain our results.  The lifetime obtained is longer than
previously estimated.  

We find interactions with the surrounding heat
bath are dominated by $\pi\pi$ scattering leading to
$\tau=8.9$~fm/c. 
Even with large nucleon densities, such as those that occur at the SPS,
inclusion of $\pi N$ scattering reduces the half-life by at most
$20\%$.  
Therefore models that only account for pion degrees of freedom
should still give reasonable results.
However, the factor of three estimated enhancement of thermal
distributions assumed at the SPS is what most dramatically effects the
half-life, reducing it to about $3$-$4$~fm/c.  
This is probably not long enough to observe DCCs in existing hadronic data.

The decay of small DCC domains is accelerated by the contribution of
$P$-wave $\pi\pi$ scattering as well as
$\pi\rho$ scattering away from threshold, decreasing
the average lifetime.
With large domains formed in the absence of nucleons, as
could occur at future colliders, the detection of DCCs looks 
more promising.

\acknowledgments

J.S. would like to thank R.~J.~Furnstahl and E.~Braaten for useful
discussions. 
J.S.  was supported by the National Science Foundation
under Grants No.\ PHY--9511923 and PHY--9258270. V.K. was supported
by the Director, 
Office of Energy Research, Office of High Energy and Nuclear Physics, 
Division of Nuclear Physics, and by the Office of Basic Energy
Sciences, Division of Nuclear Sciences, of the U.S. Department of Energy 
under Contract No. DE-AC03-76SF00098.

\end{document}